# Activities, Context and Ubiquitous Computing

**Paul Prekop and Mark Burnett**
Command and Control Division
DSTO C3 Research Centre
Fern Hill Park
Department of Defence
Canberra ACT 2600 Australia
{paul.prekop, mark.burnett}@dsto.defence.gov.au

**ABSTRACT:** Context and context-awareness provides computing environments with the ability to usefully adapt the services or information they provide. It is the ability to implicitly sense and automatically derive the user needs that separates context-aware applications from traditionally designed applications, and this makes them more attentive, responsive, and aware of their user's identity, and their user's environment. This paper argues that context-aware applications capable of supporting complex, cognitive activities can be built from a model of context called Activity-Centric Context. A conceptual model of Activity-Centric context is presented. The model is illustrated via a detailed example.

**KEYWORDS:** Activity-Centric Context**,** Context, Context-Awareness, Smart Rooms, Ubiquitous Computing

## 1. Introduction

Weiser's [43] vision of ubiquitous computing, with its invisible yet attentive computing environment, providing the right information to the right person at the right time, is an exciting vision of how to evolve computer technology. Within a ubiquitous computing environment, the computing elements and their inter-communication are largely hidden from the user, and the technology is not readily visible – it is worn or embedded in building infrastructure – and is spoken with and related to. Smart rooms, which can be seen as intense ubiquitous computing environments, are a step toward Weiser's [43] vision. Currently, there are a number of smart rooms already in use in research organizations, for example MIT's Intelligent Room [24], Stanford's iRoom Project [39], NIST's Smart Space Lab [26] and Georgia Tech's Aware Home Project [40].

While considerable work has been done in creating the physical computing devices needed to fulfil Weiser's [43] vision, little work has been done in devising ways to make the devices smarter, more attentive, responsive, and aware of their user, and their user's environment. Context and context-awareness may provide a ubiquitous computing environment with the ability to adapt the services or information it provides by implicitly deriving the user's needs from the context that surrounds the user at any point in time. It is the ability to implicitly sense and automatically derive the user's needs that separates context-aware applications from more traditionally designed applications.

Context is here defined as any information that can be used to characterize the situation of an entity. An entity is defined as the person, place, or object that is considered relevant to the interaction between a user and an application, including the user and applications themselves. A context aware application is defined as one that uses the context of an entity to modify its behaviour to best meet the context of the user [10]. Currently most attempts to use context-awareness within ubiquitous computing environments have centred on the physical elements of the environment, the user, or the user's device. While many authors acknowledge the importance of capturing the cognitive elements of a user's context [34], little work has been done to develop models to support this.

This paper initially reviews previous approaches to capturing and using context within ubiquitous computing environments. Next a conceptual model of context is presented. This model differs from many of the previous approaches, because it focuses on supporting a user's cognitive activities, rather then supporting the user's movement through a physical environment. The conceptual model of context is then explored through a simple, but non-trivial example. The final section of this paper outlines our intended approach to realise this model in an actual ubiquitous computing environment.

## 2. Existing Context-Awareness Approaches

One of the earliest descriptions of context and context-aware applications was proposed by [33] and was based on their experiences with the PARCTAB system [32, 42]. They describe context-aware applications, as applications that are able to sense their physical and computational environment – the current location of the user, the relative location of other people, hosts, accessible devices, lighting levels, noise levels, network connectivity, social situation, and so on, as well as changes to these elements over time. A key construct they used is that actions are dependent on location. The actions performed in a kitchen, for example, are different to the actions performed in a library [33]. The PARCTAB system exploits this idea by contextutalising the functions it provides to the current location. For example, a print command will only be available in a location with an accessible printing device and will default to the printing device accessible within that location.

The use of location as the core element of context has been successfully used in a number of context-aware applications. Virtual Tour Guide is based on of stick-e notes, which use location as a trigger for the retrieval of information. Stick-e notes are related to a physical location via GPS coordinates. When a user enters the area delimited by the GPS coordinates, a short note is displayed on the user's device[2]. Similar applications are described by [5] and [28]. Hewlett-Packard's cool town project [15] extends the idea of context based on physical location, by marking up the physical space with a mobile WWW infrastructure. URLs are associated with physical locations or objects – for example exhibits in a museum. As a user with an enabled device moves through the physical space, the WWW pages associated physical locations or objects, for example the description of the exhibit, as displayed on the device [19]. Similar work is described by [36].

However, location is not the only element of the physical environment that can be used within a context-aware application. Several authors, including [6, 41, 34, 20] have explored using location, as well as a variety of different sensors, such as light, sound, movement, touch, temperature, air pressure, and so on, to build an understanding of a user's physical context. This allows the context-aware application to build up a richer understanding of the user's actions within the computing environment [18, 7, 25]. Recent work developing Java programming libraries for sensor architectures[11], and in developing sensor fusion architectures [35, 6], has begun to create the programmatic and conceptual framework needed to incorporate complex sensors architectures into context-aware applications.

Most of the context-aware applications described so far tend to focus on the *external* dimension of context, the elements of the physical environment, including location, proximity to other objects, temperature, time, lighting levels, and so on [14]. While the external dimension of context is important, and as the applications described above illustrate, very useful for context-aware applications, it is not the only dimension of context that needs to be captured and used [22]. To extend context-aware application into more cognitive domains, such as information retrieval, decision making, situation monitoring, product design, and so on, the *internal* dimension of context – users goals, tasks, work context, business processes, personal events, communication, emotional and physical state – also needs to be captured [14].

Several authors have proposed models to capture the internal elements of context. For example [34] propose a hierarchal structure to codify both the internal and external elements of context. At the top level, the hierarchy includes Human Factors, which includes users, social environment, activity, and Physical Environment, which includes conditions, infrastructure, locations, and so on. Within each category, relevant features can be identified whose values determine the user's context. The idea of this hierarchy is that it provides a structure to capture the relevant elements of context.

Several context-aware applications capable of deriving a user's information needs have been developed. The Watson project [3] and IntelliZap project [13], use the user's information context, in the form of open web pages, and opened documents, as a guide to web and database searching. Similar work has been described by [21], and The *Remembrance Agent (RA)* [31] is similar again, and proactively provides relevant information to a user working with Emacs. A similar project is described by [16]. Hybrid approaches that attempt to take account of elements of both the internal and external dimensions of context are in the early stages of research. For example, the Context-Aware Internet Access project is investigating the utility of representing a user's context in accessing a web site as five elements along three axes – Personal (User, Task), Technological (Device) and Environmental (Location, Time) [4].

Currently most of the existing approaches to capturing and using context within ubiquitous computing environment and smart rooms have focused on the user's external context. While, this has resulted in some interesting, and useful applications, these approaches generally don't provide support for more cognitive activities, for example, the activities involved in systems design, or the activities involved in organising a workshop, or writing a paper, and so on. To address this limitation, we propose a model of context that focuses on capturing and using the context that surrounds the performance of an activity by an agent. This model is described in the next section.

## 3. Activity-Centric Context

This section discusses a slightly different view of context, the activity-centric view, which focuses on the context that surrounds the performance of an activity by an agent. This view of context may be more suitable for the development of context-aware applications to support complex cognitive activities.

Contexts have been defined as complex, rich objects [23] that contain information relevant to the problem or domain being examined [1]. This information is used to characterize the situation of an entity, where an entity is a person, place, or object [10]. However, simply defining context isn't enough to be able to use the concept of context to develop context-aware applications. It is also important to understand the properties of context and the relationships between context and other closely related concepts, especially, tasks or activity and users or agents [37].

The focus of the activity-centric view of context is on the information that surrounds the performance of an activity by an agent. The key components of the activity-centric view of context are agents and activities. An agent can be a single person, a group of people, or an intelligent (or semi intelligent) machine. The agent is any entity performing the activity. An activity is a description of something being done by the agent. An activity can be as broad and loosely defined as, *working for DSTO*, or as precisely defined as *get a cup of coffee*. The scope of the activity is defined by its use. As will be shown later, when discussing cascading contexts, this variability of scope of activity is an important element of context.

In terms of previous definition of context [10], the situation or entity we are considering is the performance relationship that links a particular agent, to a particular activity at a point in time. This is shown in Figure 1.

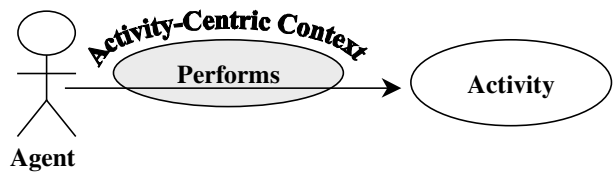

*Figure 1: Elements of Activity-Centric Context.*

Since the context being considered surrounds the performance of an activity by an agent, the context will only come into existence when the activity is being performed. The context is created at the start of an activity, exists for and is used throughout the performance of the activity by the agent, and ends when the activity being performed ends [9, 38]. While Figure 1 shows only a single agent performing a single activity, a more realistic description would show multiple agents performing the same activity, or have one agent moving between performances of several different activities.

Activities vary in scope, from the very broad to the very specific, with broad activities often containing more refined or specific activities. For example, the broad activity of *working for DSTO* would include more refined activities like *working on the pervasive computing task* which itself may be refined by *exploring the role of context in pervasive computing*, which may be refined into even more specific activities. Since the performance of each activity by the agent creates a context, the performance of a collection of nested activities (like those described above) would result in a collection of nested or cascading contexts. This is shown in Figure 2.

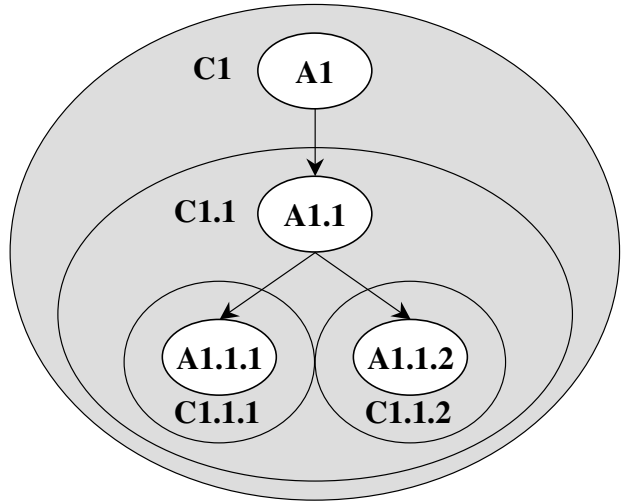

*Figure 2: Cascading Activities and Contexts.*

Activity A1, for example, can be refined by activity A1.1, and activity A1.1 and be refined by A1.1.1, and A1.1.2. The context surrounding all four activities can be described as a cascade downwards with the context C1, surrounding activity A1, as well as all the sub-activities associated with activity A1. The same is true for context C1.1, which surrounding activity A1.1, as well as all A1.1's sub-activities. The bottom level contexts C1.1.1 and C1.1.2, only surround the bottom two sub-activities, A1.1.1, and A1.1.2. Important from the perspective of activity-centric context, is the influence the various contexts have on the activities. In Figure 2, Context C1, influences, not only Activity A1 but also influences all the sub-activities – A1.1, A1.1.1 and A1.1.2. As will be shown later, the cascading effect of this influence is an important part of the activity-centric view of context.

As discussed previously, context can be seen as a container, holding information relevant to the problem or domain being examined [1]. However, the major problem with modelling context from any perspective is identifying *a priori* the information that exists within the context [12]. In the case of the activity-centric view of context, the problem is how do you identify the information relevant to the activity being performed, before the activity has been performed?

We approach this problem in two ways. The first is to identify generic information elements that would be captured by the context surrounding any kind of activity, regardless of the nature of the activity. For an activity-centric view of context, these generic information elements would include the resources needed by the activity and the process for applying them. Resources describe anything used to perform the activity, including information, computing devices or applications, other people, and so on. Process describes how the activity could be performed, and how the resources are applied to achieve the over arching goal of the activity. As will be discussed later, there are potentially other information elements that may need to be captured by context.

The second approach to dealing the problem of identifying the contextual information relevant to the activity being performed is to build all specific contexts from more generic contexts. Within most complex domains, it is often possible to identify generic types or classes of activities [17, 30] that capture the specific activities that would be performed within the domain. For each of these generic types or classes of activities, generic contexts also exist. When a specific instance of a generic activity is performed, the context surrounding the specific instance of the activity can be derived from the generic context surrounding the generic, parent activity. This relationship is captured in Figure 3.

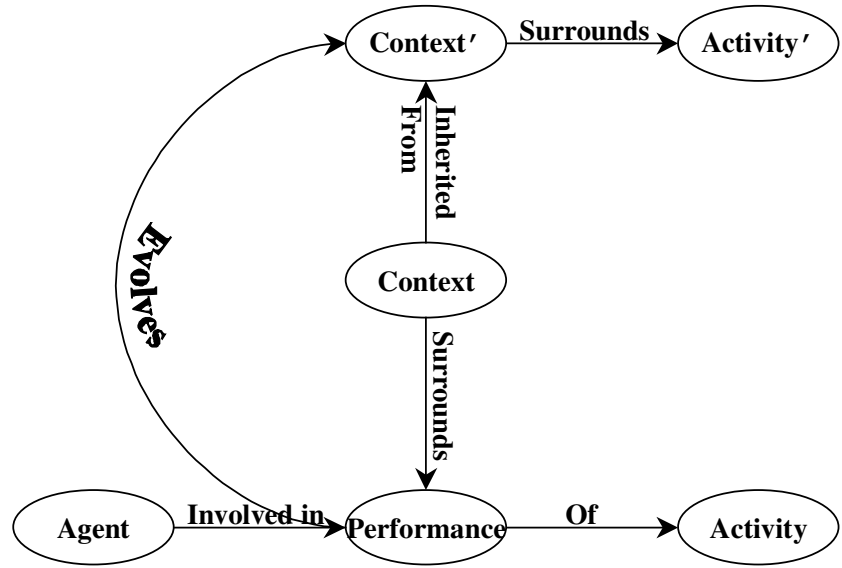

*Figure 3: Elements and Relationships within Activity-Centric Context.*

An agent is involved in the performance of an activity. The performance of an activity is surrounded by a context. The context surrounding the activity is specialised from a higher level, more generic context, context', which surrounds a higher-level more generic activity, activity'. The generic contexts and activities (context', and activity') can be modelled a priori for various domains. For example a smart room or ubiquitous computing environment that implemented the activity-centric view of context, could monitor the agent's performance of an activity, and together with machine learning techniques evolve context' to better represent the context that surrounds the particular class of activities.

As well as several agents performing the one activity, one agent may be involved in more then one activity, and context simultaneously. While an agent may be involved in simultaneous activities, the agent will only ever be explicitly focusing on one particular activity, and hence one context at any given time. The context the agent is explicitly focused on is called the agent's *contextual focus*. While an agent may only have one contextual focus, the contexts surrounding the agent's other activities continue to exist. Depending on the implementation of the smart room or ubiquitous computing environment, and the nature of the activity and context, the context may simply exist in suspended state, waiting for the agents focus to return, or the context may continue to in some way influence activities.

Since context surrounds the performance of a purposeful activity, we also argue that any artefacts resulting from the activity must in some way embed the context that exists when the activity was performed. For example, a system design activity produces a collection of system design documents – database schema, class diagrams, use-cases, and so on. Surrounding the performance of the system design activity is the system design context, created by the agents involved in the performance of the systems design activity. The system design context contains the conceptualisation of the world, descriptions of the problem being solved, the assumptions made by the design team, and so on. These elements of the system design context will be tacitly embedded in the design documents produced.

From a knowledge management perspective, understanding and exploiting the embedded nature of context is important for providing support for the hand over of artefacts from one individual or a group to another. For example, when handing over a system design from the design team to the development team, it is not only necessary to pass the formal documentation (the artefacts of the design activity), but it is also important to pass the tacit system design context, the mental models, assumptions, and other factors, that existed when the design was developed.

### 3.1 Activity-Centric Context Example

To explore the concepts of the activity-centric view of context, this section describes how the activity-centric view of context relates to a simple activity:

> Organise a two-day workshop to meet with representatives of the various defence organisations to discuss how to apply handheld computers within their organisations.

The organising a workshop activity involves many sub-activities, including selecting and booking flights, hotels, organising meetings, presentations and technology demonstrations. As this activity is being performed, the agent performing it creates an *Organising Handheld Demonstration Workshop* context, which will only exist while the agent is organising this workshop. As discussed previously, activities and the contexts that surround them don't exist in isolation, but rather cascade, following the structure of activities from the general to the specific. The activity of organising the workshop exists within some wider context, of a task or a project. This task or project context doesn't exist in isolation either, but exists within some kind of wider context, perhaps some kind of work, or job context. Figure 4, shows the context surrounding the performance of the organising a workshop activity by the agent. The context surrounding this activity is in turn surrounded by the task context, which is in turn surrounded by the project and job contexts.

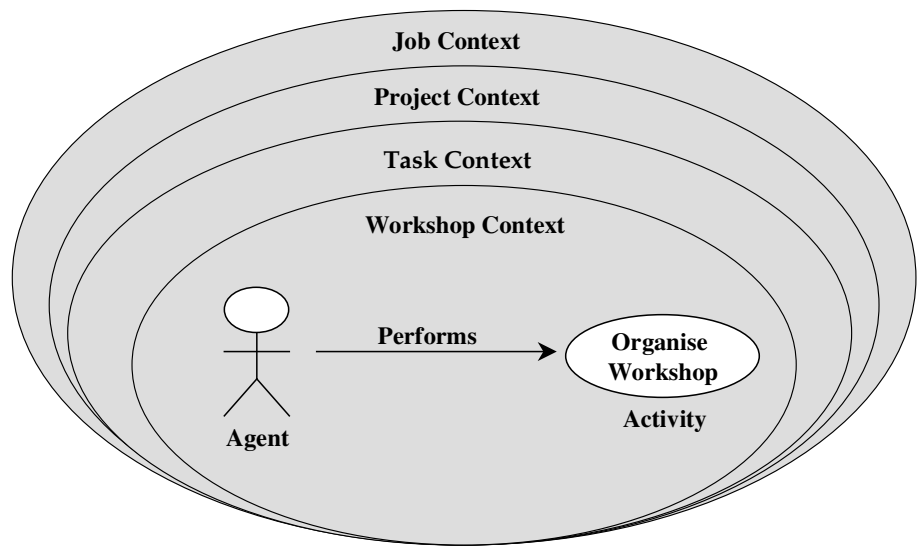

*Figure 4: Context Surrounding Organising a Workshop Activity.*

The nature of the contexts, and the activities they surround are described in detail, in Table 1.

| Activity | Context | Description |
| --- | --- | --- |
| Work as a researcher within DSTO. | Work Context. | Top-level activity, from which all job related work is derived. In this example, the Job Activity would be *Work in DSTO*<br><br>Activity, and context exists for as long as employed in job |
| Work on the Ubiquitous computing and smart room project. | Project Context. | Top-level activity that describes current focus of work. In this example, the project activity would be *Research Smart Rooms and Ubiquitous Computing*<br><br>Will change slowly overtime.<br><br>May have several project activities in parallel, or may perform other professional activities in parallel with Project activity. |
| Technology transfer of handheld computing devices into defence organisations. | Task Context. | Sub-activity of the project activity. There would be many parallel sub-activity under taken as part of the *Research Smart Rooms and Ubiquitous computing* activity.<br><br>May change rapidly, and may rapidly need to switch between the different parallel activities.<br><br>In this case, the sub-activity may be *Devise Application for PDA within Defence.* |
| Organise a technology demonstration of handheld computing devices for defence customers. | Organising a Workshop. | The actual activity of organising the trip. Includes organisation of meetings and presentations. Coordinating meeting facilities bookings. Car hire. Airline bookings. Hotel accommodation, etc.<br><br>Activities at this level may change very frequently, and have a short length.<br><br>Many activities at this level, and very rapid switching between them. |

*Table 1: Cascading Contexts for the Organising a trip activity*

The context that surrounds the organising a workshop activity would hold descriptions of the resources and processes needed by this activity. Resources would include information relevant to the activity, including descriptive information, such as flight schedules, hotel locations, meeting room capabilities, names and roles of contacts within the various organisations, descriptions of the various organisations, and so on, as well as the computer applications needed to perform this activity, for example an airline scheduling tool. The processes for completing this activity would include the process for booking a trip within the organisation, or the process for checking for available funds, and so on. Figure 5 shows a simple representation of the context.

```
Parent Context:
   Organising a Workshop

Agents Involved:
   Self

Resources:
   calendar, email, names, addresses, travel department, office
   applications, and so on.

Process:
   begin{
       Initial Agenda
       If not Approval of Agenda then begin
       Contact Participants
       Book Rooms
       Book Travel
   }End
```

*Figure 5: Example Representation of a Context.*

When performing the *organising the workshop* activity, the agent's contextual focus is clearly on organising the workshop. However, there are likely to be other activities that demand the agent's attention. For example, organising the repair of a broken PDA, or enhancing an existing time management tool on the PDA, and so on. Each of these different activities will have with it a related context, which has associated with it a collection of Resources and Processes. As the agent's contextual focus changes, so would the collection of Resources and Processes the agent would need to access.

For a context-aware application to be truly context-aware, it needs to understand not only the context surrounding the actual activity of organising the workshop, but also the cascading contexts, (in this case Work or Job Context, Project Context and Task Context), that also affect the organising the workshop activity, because each of these contexts provide vital information, in the form of facts, or processes, that are needed to successfully perform the organising a workshop activity. For example, the process needed to gain approval for the workshop might require getting approval from a supervisor, as well as approval from the person in charge of monitoring funds. The process to do this would be an organisation wide process, and likely to be contained in the Work Context (however, each work-place is likely to have a very different process). The information describing who is the approving supervisor may be contained within the Project Context. The culture and approach used to justify the workshop may be contained within the Organisational Context, or it may be contained within the Project Context.

Since an agent can be involved in any number of different activities, and hence be affected by different cascading contexts, the information provided by the different contexts will change. For example, as well as being involved in a *Research Smart Rooms and Ubiquitous Computing* project, the agent may also be involved in a professional activity, for example, sitting on job interview panels, and recruiting new staff. This may involve organising a two-day workshop at a University to talk with potential recruits. While the process of gaining approval for this workshop (drawn from the Work Context) may be the same as for the organising a workshop activity to demonstrate handheld devices, the supervisor approving the trip could be different. The approach use to justify the trip would also be different, and so on.

As Figure 4 illustrates, any context-aware application used to support the cognitive activity of organising a workshop needs to draw on information from all the contexts that surround that activity. A context-aware application that simply focuses on the process of organising a workshop would either fail to have all the information, or wouldn't be able to distinguish between organising a workshop to demonstrate the use of handheld devices and organising a workshop to recruit staff. Also technology focused on recording preferences, and previous interactions, (as described by [22]) would also fail, unless it was able to somehow distinguish what set of preferences to use. For example, should technology use the set of preferences or previous interaction from organising a workshop to demonstrate handheld devices, or should it use the set of preferences or previous interactions used to organise a workshop to recruit staff? Unless it is able to tell difference, it will only have a 50% chance of getting the preferences right.

## 4. Developing Context-Aware Applications

The computational environment needed to support the activity-centric view of context could be some kind of smart room, ubiquitous computing, or even desktop computing environment. The main requirement of the environment would be that most of the agent's activities are performed within the environment, and that the environment is capable of capturing the agent's behaviour. The applications that make up the environment must also be context-aware and be able to modify their behaviour depending on the agent's activity-centric context.

A key component of any environment used to capture and activity-centric context would be some kind of Context Manager (CM). The CM would act as the interface between the agent's activities and context-aware applications used by the agent to do its work. A possible sequence of activities between the agent, the Context Manager (CM) and the computational environment used to support the activity-centric view of context, is shown in Figure 6.

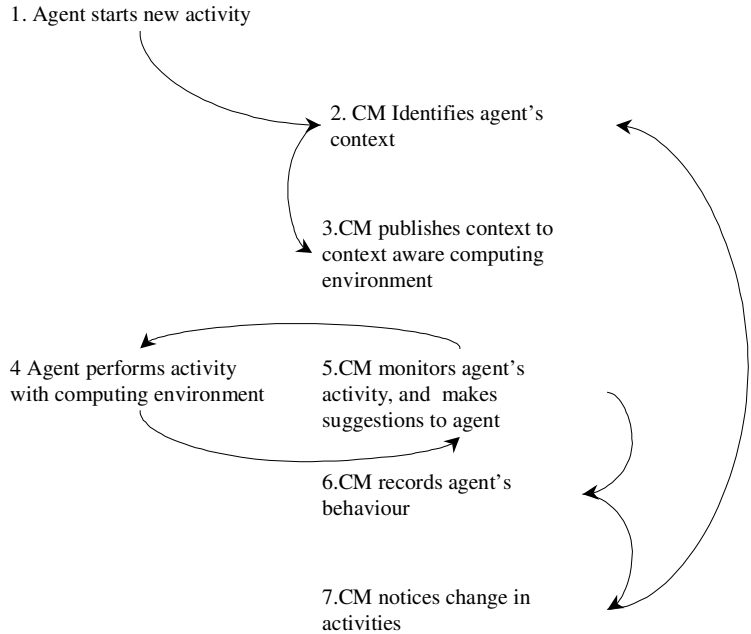

*Figure 6: Interaction between the user and environment.*

The sequence begins when the agent starts a new activity (Stage 1). The Context Manager (CM) identifies the context that surrounds this activity (Stage 2). The CM then publishes the agent's context, and contextual hierarchy, together with other global context information to all the context-aware applications within the environment (Stage 3).

The agent performs activities using the context-aware applications, which use the published context, global information, and perhaps their own local context information to support the agent (Stage 4). The CM monitors the activities being performed by the agent, and may make suggestions to the agent (Stage 5).

The CM also records the agent's behaviour (Stage 6) to help evolve what it knows about the different contexts. Once the CM detect that the agent's behaviours is moving outside the scope of the current context (Stage 7), it assumes a new activity is being performed by the agent (Stage 1), and again attempts to identify the agent's context (Stage 2).

One of the interesting questions raised by the above description of the activity-centric context-aware environment is how is the agent's context identified? (Step 2, Figure 6) Context could be identified via the location of the activity, the time of the activity, or the unique sequence micro-function performed by the agent. Recent work by [25, 29] has shown that many activities do have elements of prediction to them, and can be predicted based on patterns of actions over time, [25], or at a micro-level, by the sequence of actions the agent performs [29]. Context can then be indexed by the unique pattern of activities performed by the agent. Depending on the activity and the environment in which it is implemented, this may be as detailed as the low level collection of key strokes and mouse movements, through to higher level actions such as sending e-mail, or browsing specific WWW pages.

A context-aware environment may also seek confirmation from the agent as to their contextual focus, and use this feedback to customise the environment to support the associated activities. For example the environment detects a number of cues and on the basis of them and information about higher-level contexts derives that the agent is engaged in, say, organising a workshop, it may then seek confirmation of this.

A second important implementation issue would be the knowledge representation methods used. As discussed previously, a context contains Resources and Processes for applying them, and are specialised from more generic contexts. The representation of context requires techniques that can cope with a variety of data types, as well as support specialisation. Currently, we are considering frames as one potential approach for representing contexts [8].

## 5. Conclusions and Future Work

This paper has presented a conceptual model of context, based around an agent's performance of an activity. This view of contexts differs from many of the previous approaches, because it focuses on creating context-aware applications that support cognitive activities, rather then context-aware applications that focus on time, location or other elements of the physical environment.

The key elements and processes within the activity-centric view of context are summarised in Figure 3. An agent is involved in the performance of an activity. The performance of an activity is surrounded by a context. That context is specialised from a higher level, more generic context, which surrounds a higher-level more generic activity. A key function of the activity-centric view of context is the evolution of the more generic contexts over time.

The utility of the activity-centric view of context was explored through a small real work example – organising a 2-day workshop. A brief, high-level approach to for developing an environment to support the activity-centric view of context was also described, together with some of the key technical issues, and possible ways to address them.

While this paper has focused on exploring the utility of the activity-centric view of context to support context-aware applications, the activity-centric view of context may also be useful within the area of knowledge management.

As discussed previously, an activity-centric context is created when an activity is performed. Any artefact resulting from that activity, for example, design documents, or workshop agenda, and so on, can be seen as being surrounded by the context, (perhaps in the form of the conceptualisations, assumptions, rules, processes, facts, and so on) that existed when the activity was performed. From a knowledge management perspective, capturing and exploiting the context that surrounds an artefact may be one potential way to capture the tacit understanding [27] that existed when the artefact was created.

The model of activity-centric context described in this paper, is a first attempt to unravel the complexities of the context that surrounds the performance of complex cognitive activities. There are also several conceptual issues that need to be resolved; the first is the role of goals or intents within the contexts. Currently, the goal or intent of the activity is implicit in the nature of the activity itself. However, is the model of activity-centric context improved by explicitly capturing the goal or intent of the activity? There are also several technical issues that need to be explored. These include, the nature of the information that needs to be captured by the context, as well as methods for representing and reasoning with context, and how to evolve models over time

# References

[1] M. Benerecetti, P. Bouquet, and C. Ghidini, "Contextual Reasoning Distilled," *Journal of Experimental & Theoretical Artificial Intelligence*, vol. 12, pp. 279-305, 2000.
[2] P. J. Brown, J. D. Bovey, and X. Chen, "Context-Aware Applications: From the laboratory to the marketplace," *IEEE Personal Communications*, pp. 58-64, 1997.
[3] J. Budzik and K. J. Hammond, "User Interactions With Everyday Applications as Context for Just-in-time Information Access," presented at Proceedings of the 2000 International Conference on Intelligent User Interfaces, New Orleans, Louisiana, USA, 2000.
[4] M. Burnett and C. Chapman, "Context-Aware Internet Access," presented at The 6th Australasian Document Computing Symposium, Coffs Harbour, Australia, 2001.
[5] J. Burrell and G. K. Gay, "E-graffiti: Evaluating Real World use of a Context-Aware Systems," *Interacting with Computers*, In Press.
[6] P. Castro and R. Muntz, "Managing Context Data for Smart Spaces," *IEEE Personal Communications*, vol. 7, pp. 44-46, 2000.
[7] M. H. Coen, "Design Principles for Intelligent Environments," presented at Proceedings Fifteenth National Conference on Artificial Intelligence (AAAI-98) Tenth Conference on Innovative Applications of Artificial Intelligence, Menlo, CA, USA., 1998.
[8] R. Davis, H. Shrobe, and P. Szolovits, "What is Knowledge Representation?" *AI Magazine*, vol. 14, pp. 17-33, 1993.
[9] B. Dervin, "Given a Context by any other name: Methodological tools for taming the unruly beast," presented at ISIC 96: Information Seeking in Context, Tampere, Finland, 1996.
[10] A. K. Dey and G. D. Abowd, "Towards a Better Understanding of Context and Context-Awareness," College of Computing, Georgia Institute of Technology, Atlanta GA USA, Technical Report GIT-GVU-99-22, 1999.
[11] A. K. Dey, G. D. Abowd, and D. Salber, "A Context-Based Infrastructure for Smart Environments," presented at the 1st International Workshop on Managing Interactions in Smart Environments (MANSE '99), Dublin, Ireland., 1999.
[12] B. Edmonds, "The Pragmatic Roots of Context," presented at Modelling and Using Context. Second International and Interdisciplinary Conference (CONTEXT'99), Trento, Italy, 1999.
[13] L. Finkelstein, E. Gabrilovich, Y. Matias, E. Rivlin, Z. Solan, G. Wolfman, and E. Ruppin, "Placing Search in Context: The Concept Revisited," presented at Tenth International World Wide Web Conference (WWW10), Hong Kong, 2001.
[14] J. Gwizdka, "What's in the Context?" presented at Computer Human Interaction 2000 (CHI2000) – Workshop The What, Who, Where, Why and How of Context-Awareness, The Hague, The Netherlands, 2000.
[15] Hewlett-Packard's Cool Town, Web Site: http://cooltown.hp.com/
[16] J. I. Hong and J. A. Landay, "A Context/Communication Information Agent," *Personal and Ubiquitous Computing*, vol. 5, 2001.
[17] S. Kethers and M. Schoop, "Reassessment of the Action Workflow Approach: Empirical Results," presented at Fifth International Workshop on the Language-Action Perspective on Communication Modelling (LAP2000), Aachen, Germany, 2000.
[18] C. D. Kidd, R. Orr, G. D. Abowd, C. G. Atkeson, I. A. Essa, B. MacIntyre, E. Mynatt, T. E. Starner, and W. Newstetter, "The Aware Home: A Living Laboratory for Ubiquitous Computing Research," presented at The Second International Workshop on Cooperative Buildings (coBuild'99), 1999.
[19] T. Kindberg, J. Barton, J. Morgan, G. Becker, D. Caswell, P. Debaty, G. Gopall, M. Frid, V. Krishnan, H. Morris, J. Schettino, B. Serra, and M. Spasojevic, "People, places, things:

Web presence for the real world," presented at Third IEEE Workshop on Mobile Computing Systems and Applications, Monterey, CA, USA, 2000.
[20]     E. Kovacs, K. Rohrle, and B. Schiemann, "Adaptive Mobile Access to Context-aware services," presented at First International Symposium on Agent Systems and Applications, and the Third International Symposium on Mobile Agents, Palm Springs, CA, USA, 1999.
[21]     S. Lawrence, "Context in Web Search," *IEEE Data Engineering Bulletin*, vol. 23, pp. 25-32, 2000.
[22]     H. Lieberman and T. Selker, "Out of Context: Computer Systems that adapt to, and learn from, context," *IBM Systems Journal*, vol. 39, pp. 617-632, 2000.
[23]     J. McCarthy and S. Buvac, "Formalizing Context (Expanded Notes)," Stanford University, Stanford, CA, USA. Technical Note STAN-CS-TN-94-13, 1994.
[24]     MIT's Intelligent Room, Web Site: www.ai.mit.edu/projects/iroom
[25]     M. C. Mozer, "An Intelligent Environment must be Adaptive," *IEEE Intelligent Systems*, March/April, 1999.
[26]     NIST's Smart Space lab, Web Site: http://www.nist.gov/smartspace/theLab
[27]     I. Nonaka and H. Takeuchi, *The Knowledge Creating Company: How Japanese Companies Create the Dynamics of Innovation*, Melbourne: Oxford University Press, 1995.
[28]     J. Pascoe, "Adding Generic Contextual Capabilities to wearable computers," presented at The Second International Symposium on Wearable Computers, Pittsburgh, USA, 1998.
[29]     A. P. Pentland, "Smart Rooms," in *Scientific American*, 1996.
[30]     P. Prekop and G. Kingston, "Implementing C4ISR Architecture Framework – An Australian Case Study," presented at 6th International Command and Control Research and technology Symposium (6th ICCRTS), Annapolis, MD, USA, 2001.
[31]     B. J. Rhodes and P. Maes, "Just-In-Time information retrieval agents," *IBM Systems Journal*, vol. 39, pp. 685-704, 2000.
[32]     B. N. Schilit, N. Adams, R. Gold, M. M. Tso, and R. Want, "The PARC Tab Mobile Computer System," presented at the Fourth Workshop on Workstation Operating Systems (WWOS-IV), Napa, CA. USA, 1993.
[33]     B. N. Schilit, N. Adams, and R. Want, "Context-Aware Computing Applications," presented at IEEE Workshop on Mobile Computing Systems and Applications, 1994.
[34]     A. Schmidt, M. Beigl, and H.-W. Gellersen, "There is More to Context than Location," *Computers and Graphics*, vol. 23, pp. 893-901, 1999.
[35]     A. Schmidt and K. V. Laerhoven, "How to Build Smart Appliances," *IEEE Personal Communication*, pp. 5-11, 2001.
[36]     A. Smailagic, D. P. Siewiorek, J. Anhalt, F. Gemperle, D. Salber, S. Weber, J. Beck, and J. Jennings, "Towards Context Aware Computing: Experiences and Lessons," *IEEE Journal on Intelligent Systems*, vol. 16, pp. 38-46, 2001.
[37]     D. H. Sonnenwald, "An Evolving Framework for Collaborative Information Exploration," presented at Computer Human Interaction (CHI98 – Workshop on Information Exploration), Los Angeles, USA, 1998.
[38]     D. H. Sonnenwald, "Evolving Perspectives of Human information Behaviour: Context, Situations, Social Networks and Information Horizons," presented at Information Seeking in Context '98, London, 1999.
[39]     Stanford's iRoom project, Web Site: http://graphics.stanford.edu/projects/iwork
[40]     Georgia Tech Aware Home, Web Site: http://www.cc.gatech.edu/fce/ahri
[41]     R. M. Turner, "Context-Mediated Behaviour for Intelligent Agents," *International Journal of Human-Computer Studies*, vol. 48, pp. 307-330, 1998.
[42]     R. Want, A. Hopper, V. Falcao, and J. Gibbons, "The Active badge Location System," *ACM Transactions on Information Systems*, vol. 10, 1992.
[43]     M. Weiser, "The Computer for the 21st Century," in *Scientific American*, 1991.